\documentclass[twocolumn,prd]{revtex4-1}
\usepackage{color}

\usepackage{graphicx}
\usepackage{bbm}
\usepackage{amsmath}

\begin{document}
\title{ A new state of dense matter in neutron stars with nucleon structure }
\author{Vikram Soni}
\email{vsoni.physics@gmail.com}
\affiliation{Centre for Theoretical Physics, Jamia Millia Islamia, New Delhi, India}

\begin{abstract}
The existence of  stars  with a large mass of 2 solar masses means that the equation of state is stiff enough to provide high enough pressure at large central densities. Previous work shows that such a stiff equation of state is possible if the ground state has nucleons as its constituents.  We find this to be so in a chiral soliton ( skyrmion ) model for a composite nucleon which has bound state quarks. The strong binding of the quarks in this composite nucleon is plausibly the origin of the nucleon-nucleon hard core. In this model we find a new state of superdense matter at high density which is a 'topological'cubic crystal of overlapping composite nucleons that are solitons with relativistic quark bound states. The quarks are frozen in a filled band of a unique state, which not an eigenstate of spin or isospin but an eigenstate of spin plus isospin, $ \vec S + \vec I = 0$. 

In this alternative model we find that all neutron stars have no regular `free'quark matter.  Neutron stars whose central density crosses a threshold  baryon density  of approximately, $n_b \sim 1/fm^3 $, will become unstable  and go through a decompression (sudden) density discontinuity to conventional quark matter. Sequentially, this contraction of the core of the star will soften the equation of state release a large amount of gravitational potential energy which can give rise to a shock wave and matter ejection. Since the merger of two neutron stars gives a compact state whose mass is larger than the allowed maximum mass, this will be followed by a jet and a short gamma ray burst while transiting into a  black hole .

\end{abstract}

\maketitle

\section{Introduction}
\label{introduction}

The object of this work is to reconcile and relate some recent observations on neutron stars and their merger in a consistent manner.

i)  
Till the recent findings of the high mass ( $\sim$ 2 solar mass ) neutron stars \cite{shapiro10,freire} , neutron stars were expected to have  nuclear matter in the outer regions of lower density giving into quark matter cores in the interior region of high density, close to their centres.
However,  such stars have a stiff, non relativistic nucleon matter exterior  pushing into a
softer relativistic quark interior - an unstable situation. In this case a star with a quark core is stable only if the nuclear matter to quark matter transition takes place in a small window at low pressure \cite{Soni1}. It is also for this reason that most neutron stars with conventional quark matter cores and in particular with meson condensates have smaller maximum masses ( ~1.6 solar mass), as has been pointed in the work above. A recent review \cite{BH} also points out that pure conventional (for example, MIT bag) quark matter stars are very unlikely to ever have an EOS as high as ,  $ M_{max}\sim 2 $ solar mass.

It is well known that there are many purely nucleon based neutron star models that have neutron stars with maximum mass slightly above 2 solar masses, for example, the APR 98 equation of state (EOS) of Akmal, Pandharipande and Ravenhall \cite{APR}.  All such equations of state have one common characteristic and that is a hard core that becomes operative at high density. In view of the foregoing, we investigate the following question; Is matter in neutron stars be entirely composed of nucleon degrees of freedom ?

	There is a plurality of equations of state (EOS) for purely nuclear matter at high density - APR, Bonn  potential, Paris potentials, Reid potential, Skyrme potentials, nuclear mean field theory, Bruckner Hartree-Fock  etc. Similarly, there are many EOS’s for quark matter - the MIT bag model, linear sigma model, NJL model, PNJL, PQCD models and many variants of these \cite{BH,APR,Lattimer,bla1,bla2,bla3,hat1,hat2,fuk,baym,weise1,weise2}. These equations of state  have not only numerical uncertainties but even more fundamental conceptual ones. None of them are testable at directly at high density. Besides, all these come with a large set of parameters, that allows a lot freedom of fitting, but without much conviction. 
	
	We refer the reader to the review of Baym et al \cite{BH} and the references therein, where an attempt is made to work out a hybrid model that can interpolate between nuclear EOS  at low density (nuclear saturation densities and above) and quark matter at high density. By appropriately including various repulsive interactions between quarks  such hybrid equations of state can be pushed to accommodate high mass ($\sim$ 2 solar mass) neutron stars. However,  these works do not explicitly use the  structure of nucleons with quark bound states.


ii) We  know that the nucleon is a composite object made of 3 valence quarks. If we could work out a ground state made up of such nucleons that has a nucleon nucleon hard core,  we also know that finally, at some threshold density, it should dissolve into quark matter. A faithful model of the composite nucleon then is obliged to reproduce these features. We shall use our knowledge of  nucleon structure and a possible  new solitonic nucleon crystal ground state to work some insight into the  high density EOS and the threshold density for the transition to quark matter.

iii) We have recent data on the remarkable merger of two neutron stars \cite{abbott1,abbott2} whose end state is  an object of rather large mass,  $\sim 2.7$ solar mass. 
Though we do not know with certainty if this object is a neutron star or a black hole, we do know that this event does produce a kilonova and a weak gamma ray burst. Such an event could be associated with a 'collapse' to a  denser object and thus  the EOS undergoes a drastic change - that is, goes soft, engineering an abrupt contraction. This can release a lot of gravitational energy that could be  responsible for the matter ejection  that is seen after the merger. It must be kept in mind that the usually accepted EOS's of neutron stars have a hard core repulsion between nucleons as the density goes up. Such a 'collapse' would mean that after a threshold density the hard core barrier between nucleons will dissolve into quark matter.

	Our attempt here is to work with an alternative model for the EOS of dense matter. We use a chiral theory which can describe quark matter at high density and also  give a very representative model for a nucleon with quark bound states\cite{Soni1}. This can account for the nucleon nucleon hard core interactions, without introducing any additional interactions or parameters, in the high density interval before nucleons give way to quark matter.

In  Section II we review the chirally symmetric mean field theory of quarks and a chiral multiplet of pions and fields which can describe both the composite  nucleon and quark matter \cite{Soni1} . Section III, addresses the first order transition, through a mixed phase, from  nuclear matter to conventional quark matter via the  Maxwell construction between the two phases using the popular  APR 98 nuclear equation of state for nuclear matter. This indicates that the phase transition from nuclear matter to quark matter occurs at densities close to the central density of  $\sim$ 2 solar mass stars. 

However, this analysis assumes point particle nucleons; it does not take account of the structure and  the quark binding inside the nucleon. In section IV,  we find  that the strong binding of quarks in the nucleon can change the nature of the phase transition and  move the the nuclear matter-quark matter transition to appreciably higher density. In section V we review earlier work pointing to a possible new 'topological' crystalline ground state of composite nucleons for dense matter. Section, VI and VII are a heuristic  attempt at writing down a solid crystal EOS for composite nucleons. In section VIII we show how in the passage to increasing density, we overcome the nucleon-nucleon hard core barrier at some threshold density and make the transition to a soft EOS of quark matter, releasing  enough gravitational energy to power a shock wave that can eject matter. We would like to state at the outset that in proposing this alternative model we shall bring into this context many earlier works that are of relevance.

\section{ The Theory}
In this work we look at nucleon structure in an effective chiral symmetric theory for the strong interactions that is QCD coupled to a
chiral sigma model. The theory thus preserves the symmetries of QCD. In
this effective theory chiral symmetry is spontaneously broken and the
degrees of freedom are constituent quarks which couple to a color singlet,
sigma and pion fields as well as gluons \cite{Soni1,Soni2}. Furthermore, since we  do not have 
exact solutions for a theory of the strong  interactions, we work in mean field theory in which in the first approximation we assume that
mean fields associated  the gluon fields are absent and perturbative QCD
effects are ignored.

\begin{eqnarray}
\L &=& - \frac{1}{4} G^a_{\mu v} G^a{\mu v}|_{color}
 - \sum {\overline{\psi}} \left( D + g_y(\sigma +
i\gamma_5 \vec \tau \vec \pi)\right) \psi\nonumber\\
&& - \frac{1}{2} (\partial
\mu \sigma)^2 - \frac{1}{2} (\partial \mu \vec \pi)^2
 - \frac{1}{2} \mu^2 (\sigma^2 + \vec \pi^2) \nonumber\\
&&- \frac{\lambda^2}{4} (\sigma^2 + \vec \pi^2)^2
 + \hbox{const}
\end{eqnarray}

The masses of the scalar (PS) and fermions follow on the
minimization of the potentials above. This minimization yields

\begin{equation}
\qquad \mu^2 = - \lambda^2 <\sigma>^2
\end{equation}

It follows that

\begin{equation}
\qquad m_{\sigma}^2 = 2\lambda^2<\sigma>^2
\end{equation}

For the vacuum of the theory the constant is adjusted to yield, $ <\sigma> = f_\pi,  <\vec\pi> = 0 $.

 The nucleon  in such a theory is a color singlet quark soliton in the skyrmion background with three valence quark bound states
\cite{NPA,BB}. The quark meson couplings are set by matching the mass of the nucleon to its experimental value and the meson 
self coupling is set from pi-pi scattering, which in turn sets the tree level sigma particle mass to be of order 800 - 850 MeV. For details we  refer the reader to ref. \cite{Soni1}.

This is one of the simplest effective  chiral symmetric theories for the strong
interactions at intermediate scale and we use this consistently to
describe, both, the composite nucleon of quark bound states and quark
matter. Later, we  attempt to look at a ground state that is a crystal composed of these skyrmion like composite quark soliton nucleons. We find that the strong binding of quarks in the nucleon could move the transition from the nuclear to the quark phase to appreciably higher density. 

To reiterate we work at the mean field level where the gluon interactions are subsumed in the color singlet sigma and pion fields they generate. Since we will be working with quark matter at high density confinement is not an issue. We could further add perturbative gluon mediated corrections but they are not expected to make an  appreciable difference.

\section{The Maxwell construction for  the nuclear matter to quark matter transition}

In this section we examine  the Maxwell construction for the first order transition from the nuclear phase to the quark matter phase using some typical equations of state for the two phases.

To describe  the purely nuclear phase we employ the tried and tested APR 98 \cite{APR} equation of state. For quark matter we use the simple effective chiral symmetric theory which has been used to describe, both, the composite nucleon of quark bound states and quark matter \cite{Soni1,Soni2}. Variationally, one of lowest energy ground states at high baryon density that we find in such chiral models is quark matter with a neutral pion condensate   \cite{dautry,kutschera+90}. The equation of state for neutron stars  for such a state has been obtained in \cite{Soni1, Soni2}

 A simple way to look at the transition from  nucleons  into quark matter
 is to plot,  $E_B$,  the energy per baryon, in the ground state of both, the quark matter
and the nuclear phases, versus $ 1/n_B$,  where $n_B$ is the baryon density.  For the
 quark matter equation of state see Fig.1 \cite{Soni1} in which the quark matter EOS
 is indicated by the solid curves and the APR \cite{APR} nucleon
 EOS by the dashed line. The  slope  of the common tangent between the
 two phases then gives the pressure at the phase transition and the
 intercept, the common baryon  chemical potential.

\begin{figure}[htb]
\centerline{\resizebox{8.5cm}{!}{\includegraphics{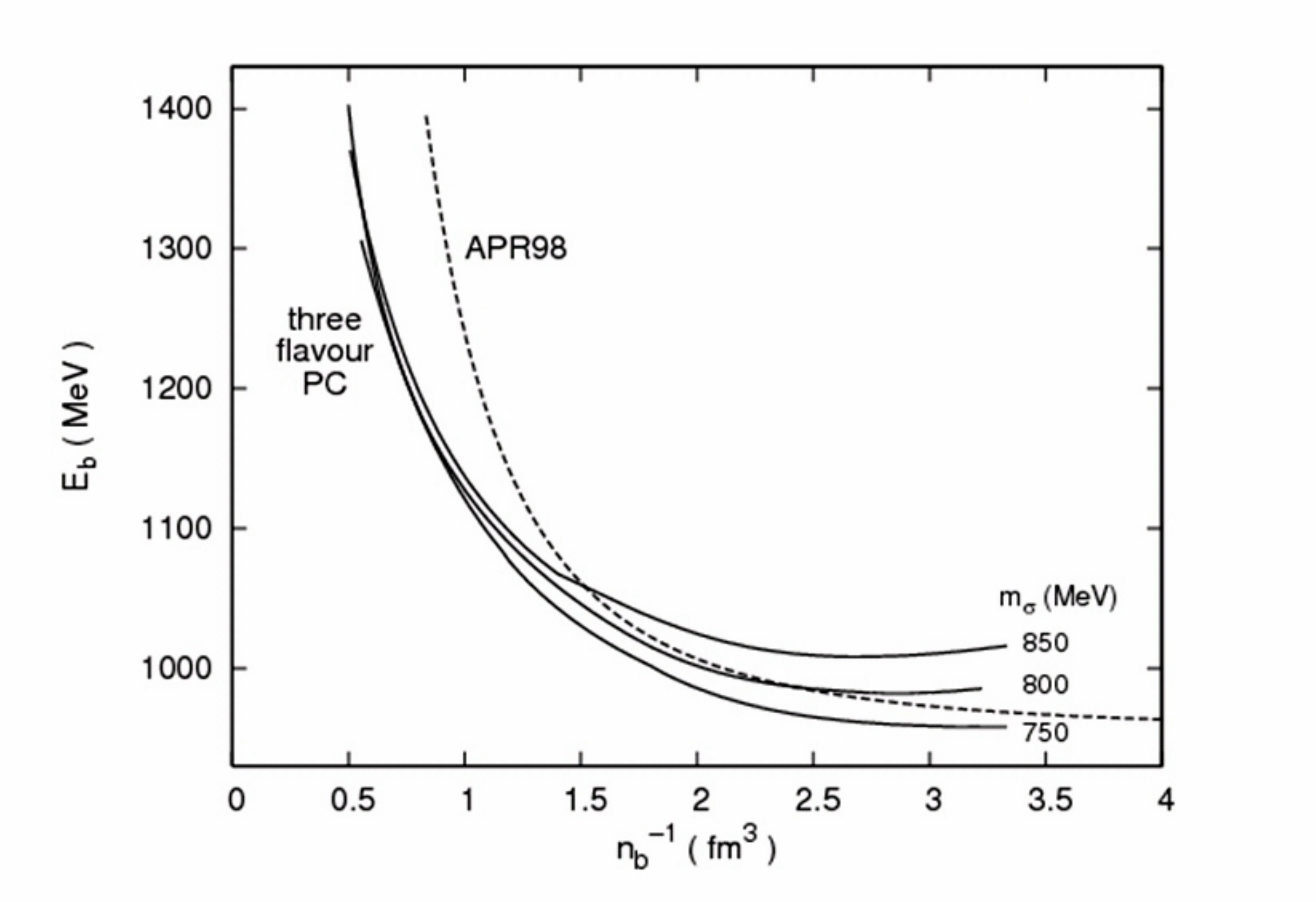}}}
\caption{The Maxwell construction: Energy per baryon plotted
against the reciprocal of the baryon number density for
APR98 equation of state (dashed line) and the 3-flavour pion-condensed
phase (PC) for three different values of  $m_\sigma$ (solid
lines). As this figure indicates, the transition pressure moves
up with increasing $m_\sigma$, and at $m_\sigma$ below $\sim$750 MeV a common tangent between these two phases cannot be obtained.\\ 
(From Fig. 2 of Soni and Bhattacharya \cite{Soni1} ) }
\label{Fig1}
\end{figure}

As can be seen from Fig.1 \cite{Soni1} , it is the tree level value of the sigma mass
that determines the intersection of the two phases; the higher the mass
the higher the density at which the transition to quark matter will take
place. In \cite{Soni1} it was found that above, $m_\sigma\sim$ 850 MeV, stars with quark matter cores become unstable as their mass goes up  beyond the allowed maximum mass.

 From Fig. 1, for the tree level value of the sigma mass $\sim 850 $ MeV,
 the common tangent in the two phases starts at  $ 1/n_B \sim  1.75$
 fm$^3$  ( $ n_B \sim 0.57/fm^3 $)   in the nuclear phase of
 APR   [A18 + dv +UIX] \cite{APR} and ends up at  $1/n_B \sim $  1.25 fm$^3$  ($n_B \sim 0.8/ fm^3$) in the quark matter phase.

In the density interval between the two phases, there is a mixed phase
at a pressure given by the slope of the common tangent and  at a
baryon chemical potential given by the intercept of the common tangent
on the vertical axis. 


Going back to the APR phase in in fig 11 of APR \cite{APR} we find that
for the APR  [A18 + dv +UIX] the central density of a star of 1.9  solar
mass is,  $ n_B \sim  0.7 /fm^3$,  which falls in the middle range
of the phase transition. On the other hand,for APR  [A18 +UIX] the central density of a star of 1.9  solar
mass is,  $ n_B \sim 0.57 /fm^3$.

Ideally we would  want the central density of the star to be a little less than
the initial density at which the above phase transition begins in the
nuclear phase. We have found that even with the simple quark matter EOS these densities are in the same ball park.

In the following, we shall present  arguments to show that the phase transition to quark matter is likely to  occur at higher density.

\section{The Nucleon}

 The above analysis assumes point particle nucleons. 
It does not take account of the structure and  the quark binding inside the nucleon
This is not captured by the Maxwell
construction. Our  attempt is to take this further by looking  at the well  accepted quark soliton (skyrmion) model  of the nucleon that is amenable to investigating quark binding properties.
We now go on to show that this could move the transition from the nuclear to the quark phase to appreciably higher density.


Following , Kahana, Ripka and Soni\cite{NPA}, we have an approximate and simple expression for the energy, $E_B$, of a color singlet nucleon soliton, with three colored bound state quarks. In accordance with the skyrmion configuration the VEV 
for the pion and sigma fields are  \cite{Soni1, Soni2}

    \begin {equation}
 <\sigma> = f_{\pi} Cos{\theta(r)}  ,  <\vec\pi> = \hat r f_{\pi}Sin{\theta(r)}  
        \end {equation}

    where, $\theta(r\rightarrow\infty) = 0$,  from the finite energy condition

and $ \theta(r\rightarrow 0)= -\pi$,  for the pion field to be well defined
     at the origin and,  $ f_{\pi}$ = 93 Mev is the pion decay constant.

The energy expression for the soliton with quark
bound states is given below.The first term below is the quark bound state(Dirac) energy
eigenvalue in the skyrmion background. In this background there is
a single valence quark bound  eigenstate of spin plus isospin, $\vec I + \vec S = 0 $,
with a color degenaracy of 3. The second term is the kinetic term from
the mesonic part. For this calculation we make the simplifying assumption,  $\sigma^2 + \vec\pi^2 = f_{\pi}^2$. In this case the potential term corresponding
to the spontaneous symmetry breaking is identically zero.
  
\begin{equation}
E/(g f_{\pi})  =  N (\frac{3.12}{X}   -  0.94) + 2\pi ( 1+ \pi^2/3) .\frac{X}{g^2}
\end{equation}

where , g is quark meson (Yukawa) coupling and N, the number of bound state quarks.
In this section we work with the dimensionless parameter, $X  =  R g f_{\pi}$, where R is the soliton
radius. This follows from a simple parameterization for radial dependence of, $ \theta(r) = \pi( r/R - 1) $, in a soluble model \cite{NPA}(see fig. 2). The
'mass' of a 'free' quark in this model is given  by,  $ m_q =   g f_{\pi}$.

Minimizing this with respect to , X 

\begin{equation}
X^2   = \frac{3.12 g^2 N}{27}\\
\end{equation}

On substitution of this value
\begin{equation}
E_{min}/(g f_{\pi})  =  2(\sqrt{\frac{3.12 N .27}{g^2} }) -  0.94N 
\end{equation}

For the nucleon soliton we must set , $N = 3$ as all three quarks sit in the bound state. 
Also, the total degeneracy of the single, $ 0^+ $, bound state is 3 - the number of colors. The soluble model above is very useful in understanding the quark bound state structure of the solitonic nucleon. However, as can be seen from Ref.\cite{NPA}(Section 6 ), compared with the soluble model an exact solution brings down the the soliton energy by close to 25 percent.
The value of the coupling ,$ g $, that fits the nucleon mass also goes down proportionately. 

In the interests of
consistency with the following section we shall choose the coupling to be, $ g \sim 7.55 $, as given in ref\cite {BGS}, which corresponds to the isolated soliton mass, $ M_{sol} \sim 976 $ Mev. 

The above formula allows us to also look at  the energy of the configuration in which two quarks sit in the bound state  and one is moved up to the continuum. Such a state will give a measure of the energy required to unbind the nucleon. 


We can easily check the possible bound states by evaluating the ratio
of  the  energy of  bound states with 2 and 3 quarks, which is given by,
$E_{min}/(N g f_{\pi})$ . 
 If the answer is less than, 1, we have a bound state, otherwise not.

\begin{eqnarray}
E_{min/}(N g f_{\pi})  &= & 2(\sqrt{\frac{27 \cdot 3.12}{Ng^2} }) -  0.94 \nonumber\\
                       &\sim& 0.464 ~\text{for}~  N = 3\nonumber\\
		       &\sim& 0.78 ~\text{for}~  N =2\nonumber\\
	        	&\sim& 1.49~\text{for}~ N =1
\end{eqnarray}

This indicates that  we have bound states only
for N = 2 and 3. Given the value of ,  $ g $,  we can  find the
energy required to unbind a quark from such a nucleon. The energy of
a two quark bound state and an unbound quark is $  2.56 g f_{\pi} \sim 1797 $ MeV  in comparison to the energy of a 3 quark bound state nucleon which is , $ 1.39 g f_{\pi}\sim 976 $ MeV.

We use the results from the parametrization for the above 'soluble' model to make some heuristic estimates below.

i) The difference between the two states above gives the binding energy of the quark in the nucleon, $ 1.17 g f_{\pi} \sim 821 $ MeV. 
	The quark binding in this model is very high. It should be noted that this is the origin of a hardcore when we bring two nucleons together. The greater the binding of quarks the greater the energy required to liberate them when we squeeze two nucleons.

ii) In this model the quark bound state eigenvalue (Fig. 2) \cite{NPA}  is  described by the figure given below.

\begin{figure}[htb]
\centerline{\resizebox{8.5cm}{!}{\includegraphics{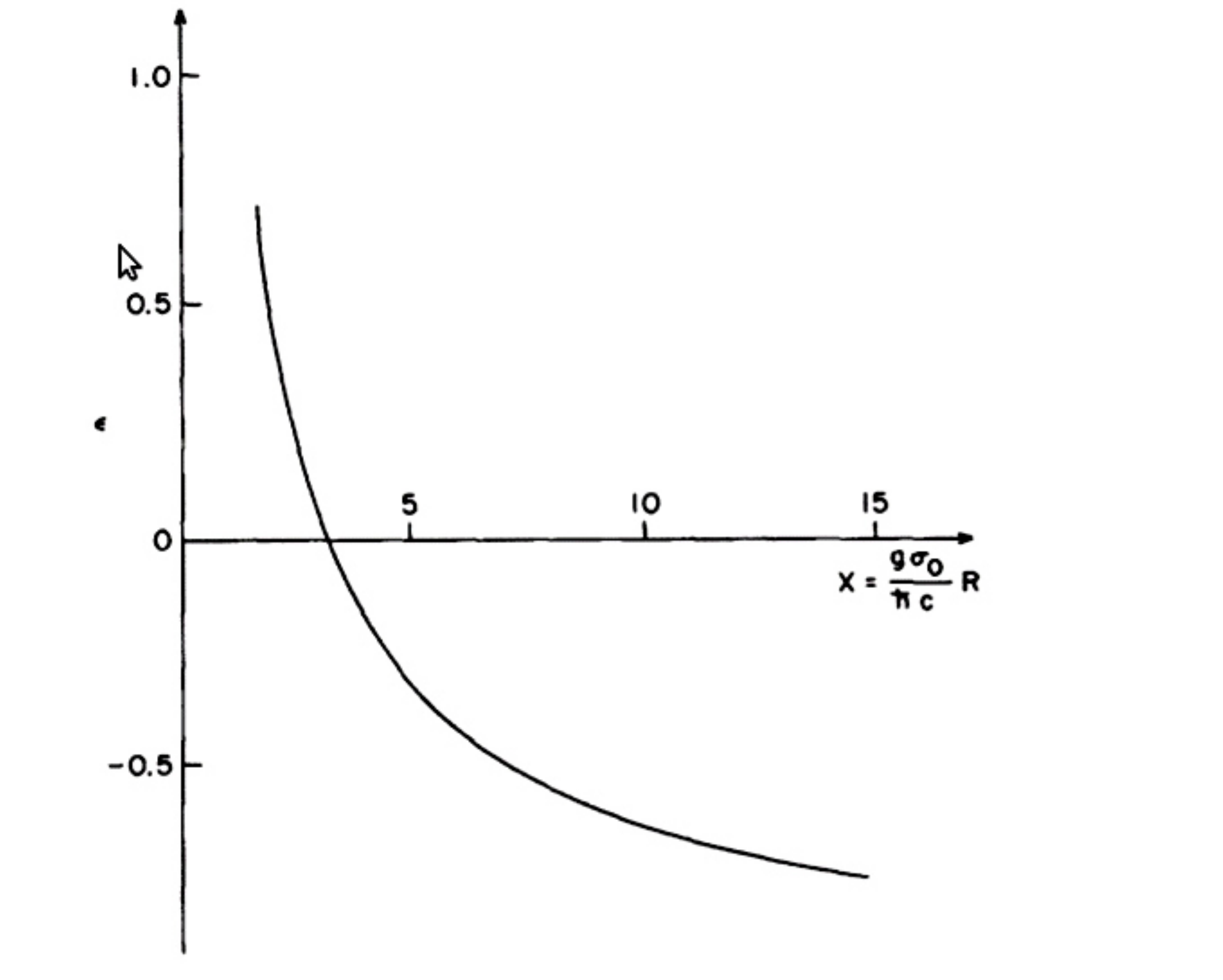}}}
\caption{Dependence of the quark energy on the soliton size $X$ in the
quark soliton model\\
 (From Fig. 2 of  Kahana, Ripka and Soni \cite{NPA}) }
\label{Fig2}
\end{figure}

 We can see that the quarks will become unbound ( go to the continuum)
when the energy eigenvalue is larger than the  unbound mass of the quark
which is given by  $m_q = g f_\pi$. This happens roughly when,  in
the dimensionless units used in Fig. 2, the energy eigenvalue, $\epsilon\sim 1.$

\begin{equation}
\epsilon\ge 1 ,   \text{at  X = 3.12/1.94 = 1.6}. 
\end{equation}
which translates into  $R \sim 0.46 fm$ for, g = 7.55 ( R, depends inversely on, g )

This is a rough  estimate of the effective radius of the squeezed nucleon at which the bound state quarks are liberated to the continuum. By assuming  that the nucleons are stacked in a cubic lattice and inverting the volume occupied by a nucleon this translates to a nucleon density of,

\begin{equation}
n_B  =  \frac{1}{(2R)^3}\sim  1.29 fm^{-3}
\end{equation}


This section has shown that the internal structure of quark binding in the nucleon not only provides the hard core but changes the nature of the phase transition that is captured by the Maxwell construction and indicates that the transition to free quark matter is likely to  be  delayed till the quark bound states meet the continiuum. More evidence of this comes from the next section.

For quark bound states in nucleons we have found above that the coupling is strong and  the binding energy is rather large $ \sim 821 $ MeV, whereas the energy scale corresponding to the inverse size of the nucleon ( $ \sim 2$ fermi) is much smaller, $ \sim 100  $ MeV. This indicates that  even when nucleons overlap the the quarks will not dissociate into a quark plasma and the nucleons get compressed but retain their identity. This is in total contrast to atomic physics, for example the hydrogen atom, where the binding energy is much smaller  than the energy scale of corresponding to the inverse size of the atom.
 
Thus the quark bound states in the nucleon may persist until a much higher
density $ n_B \sim (1 - 1.29)/fm^{3}$. In other words, nucleons can survive 
above the density range of the Maxwell phase transition  and
appreciably above the central density of the APR  2-solar-mass star. It is useful  to recall that our parametrization here is rudimentary.

\section{ The skyrme soliton composite nucleon crystal state I}

Now we move to completely different perspective on the EOS. One of
the ground states of dense nuclear matter that has been popular even
in nuclear physics, where the nucleons are assumed structureless, is
a neutron crystal \cite{PS,BaymLN}. Of course, such a ground state is
viable much beyond saturation density as nucleus matter does not show any
such tendency even for very large  nuclei. Such a crystalline state
can be also treated in a single cell Wigner Sietz approximation with
appropriate boundary conditions.

We shall first review such a calculation that was carried out by
Banerjee, Glendenning and Soni \cite{BGS} with some interesting
findings. This is a relativistic( Dirac) band structure calculation of
a cubic lattice of solitonic composite nucleons, with quarks bound in a skyrme soliton background. The quark bound state in the skyrme ( 'topological') soliton is an eigenstate of spin plus isospin, $\vec I + \vec S = 0 $, that we encountered in the last section. The relevant quark band is the $0+ $ relativistic positive parity valence band that emerges and can be tracked as function of baryon density, which  is plotted in their Fig 1 .  In the band the quark wave functions peak at the centres of the soliton. It is important to note that this state has a color degeneracy of 3 and is completely occupied and thus the 0+ band is full. Below, we highlight issues of this ground state that were not emphasized in Ref.\cite{BGS}.

There is a large gap between the top of this band  and  the next energy states which belong to the positive energy continuum. Thus quarks are frozen or fermi blocked and cannot behave as regular quark matter till this band reaches the the positive energy
continuum which happens at a radius of $\sim 0.5 $ fermi (or a cell
length of 1 fermi). For a cubic lattice this translates into a baryon density \cite{BGS} of , $1/fm^{3}$. 
(We note that this calculation uses  a coupling
constant g = 7.55, which yields a soliton mass M = 976 MeV, and an
equilibrium R ~ 1.22 fm.). 

It be seen from the figure, the band spreads out above and below the single bound state we found in the preceding section. Thus the  density at which the top  of the band meets the continuum is slightly lower than the density at which the single  bound state merges with the continuum.
As in last section, this work employs the same  approximate soluble model parametrization of the sigma and pion fields. An exact solution will yield a lower value of, g, in turn increasing the value of, R ($ R \sim 1/g $ ), and lowering the density at which band gets to the continuum.

\begin{figure}
\centerline{\resizebox{9cm}{!}{\includegraphics{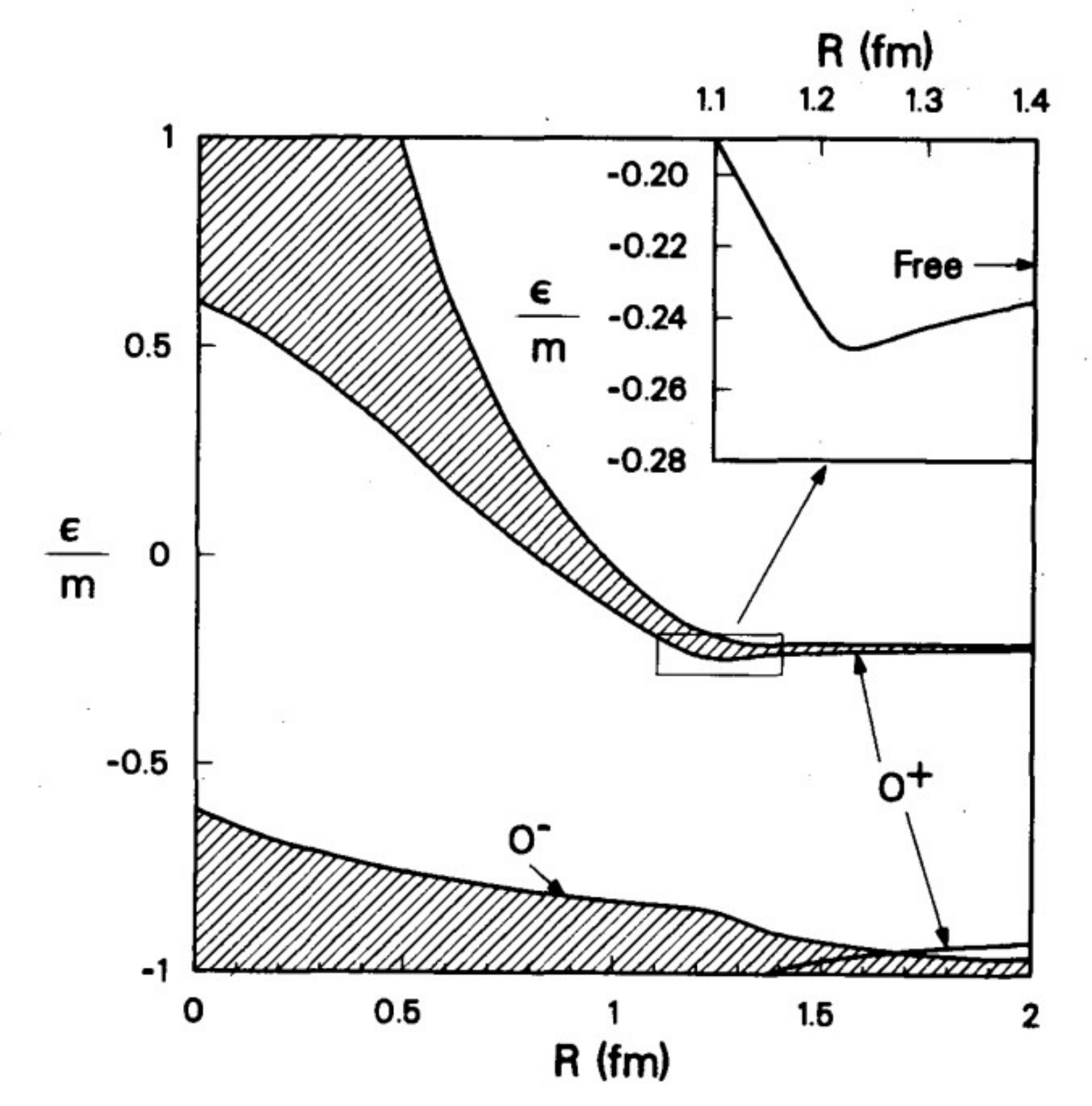}}}
\caption{Eigenvalues of the valance (0+) and sea (0-) orbitals of quarks in
soliton matter as a function of Wigner-Seitz cell radius, R.  The
band of levels that develops as the spacing decreases is shown by
the shaded region.  
( From Fig. 1 in B. Banerjee, N. Glendenning and V. Soni Physics Letters B, Volume 155, Issue 4,(1985))
}
\label{c60}
\end{figure}

This is an independent validation of the fact that in this model the onset of conventional quark matter occurs at much higher density than indicated by the Maxwell construction of the earlier section. Till this density at which the bands intersect the medium behaves as a color insulator.

\textit{ This is a new state of dense matter that is quite different from a regular, 'free' quark matter state and from conventional nuclear matter.  Such a state  is a direct consequence of our composite soliton nucleon structure.}

The quarks live in continuum relativistic Bloch states but due to the filled band and a large band gap they are blocked out. It should be noted that this work does not include the interaction between nucleons and also does not take account of the quantization of the solitons to yield  well defined nucleon / neutron states . However, this is an independent confirmation of the fact that active ( 'free') quark matter comes into play only well above the density indicated by the the Maxwell construction in Sec. III. The last two sections have established the existence of this new ground state that exists till a threshold density of approximately, $ n_B \sim (1)/fm^{3}$. The following two sections are devoted to the listing of the features of the EOS for the new ground state and a heuristic estimate of the same.


\section{ Soliton - Soliton Interaction and Quantization Energy in a pure skyrmion model}

i)   Klebanov\cite{Kleb} 
considers a cubic crystal of pure skyrmions (without quark bound states). This paper works out the most favourable spin/isopin configuration, the so called attractive 'tensor' interaction, between skyrmions. However, ref \cite{Kleb} works in the chiral limit ( $ m_{\pi} = 0$ ), whereas realistically the the tensor interaction is not long range as it is modulated by the factor, $\exp^{ (-m_{\pi}r)}/(r^3) $. This will strongly reduce the attractive tensor interaction at larger, $r$. He also estimates the energy of canonical quantization (or isorotational energy) of the whole crystal to yield states of good isospin and the third component of isospin.

Figs 1 of \cite{Kleb} calculates the classical energy per baryon
(skyrmion),$ E_1 = M_ {cl} $, which includes the 'tensor' interaction
between skyrmions, versus volume per baryon, where the \textit{free
skyrmion classical mass (864 MeV in their case)} has been subtracted. This work goes on to include the energy of canonical quantization in their Fig. 2, which calculates the sum of classical and isorotational energies, $ E_2 = M_{cl} + 1/(8\lambda_I)$ ( $ \lambda_I $ is the moment of inertia ),  per baryon versus volume per baryon, where the \textit{nucleon} \textit{mass (938 MeV)} has been subtracted. The difference between the two, $ E_2 - E_1$, then gives us the contribution of the isorotational energy of canonical quantization. We shall use these estimates in the following section.






 However, there is a caveat. As can be seen from Fig.2 in \cite{Kleb}
 the minimum energy per nucleon occurs around , $ 1/n_B \sim 4 fm^3$
 or $n_B \sim 0.25/fm^3 $. Below that density the crystal is not a stable
 state. We should therefore treat this as a variational ground state only above this density.


ii)  The Projection approach
There is point of contention here. Many authors have an alternative approach and project out good spin, isospin states, $ \vec J = \vec I = \frac{1}{2} $ corresponding to the nucleon.  The soliton is considered to be a coherent wave packet - a linear super position of all, $ \vec J = \vec I =  (n + \frac{1}{2}) $ states (see Ref\cite{Soni2} ).  The maximum weight comes from the lowest,  $ \vec J = \vec I $ states.  We may then  make the approximation that the soliton is an equal linear superposition of the nucleon, N, and $\Delta$ states and set the soliton energy to be midway between $ M_N $ and $ M_{\Delta}$.

\begin {equation}
     M_{soliton} = M_N + \frac{1}{2} (M_{\Delta}  - M_N )
\end {equation}
	or
\begin {equation}
     E_B  = M_N  =  	M_{soliton}  - \frac{1}{2} (M_{\Delta}  - M_N )
\end {equation}
	
		Unlike the former case of the isorotational energy of collective quantization, which is additive and raises the nucleon mass, in this case, the nucleon energy is well below that of the soliton. This matter is still not a settled issue. In  passing, it should also  be pointed out that in most works on solitons with quark bound states \cite{Soni2,BGS} the attractive 'tensor ' interaction between solitons has not been taken into account.

\section{ The skyrme soliton composite nucleon crystal state II }

Using the learning from the last sections, we shall make a heuristic
attempt  to write down the  EOS for a cubic crystal of composite solitons with quark bound states that we have introduced earlier with some assumptions. Given all the approximations in the previous sections, the following should be viewed as a pedagogical exercise. A more complete version is in progress and will be presented in a later work.

First we write down the energy, $E_{CS}$,  of an isolated composite
soliton. This follows from chirally symmetric linear sigma
model\cite{Soni1} used to construct the soliton with quark bound states,
where $ m_\sigma = 850$ MeV  \cite{Soni1,Soni2}. The first term below is the quark bound state  eigenvalue energy (of the soluble model), the second term is the kinetic term from the mesonic part. In contrast to Sec IV, here we relax the constraint on fixed the vacuum expectation value (VEV) for the meson fields; $  <\sigma>^2 + < \pi>^2 = F^2 $ is the sum of the square the expectation value of the sigma and pion fields which can be density dependent and therefore different to, $ F^2 = f_{\pi}^2 $. We therefore include the potential or symmetry energy term which is the last term below.

\begin{eqnarray}
E_{CS}/(f_{\pi}) &=& 3(\frac{3.12}{Y}- 0.94(g) Z) +  2\pi (Y) (Z^2)
( 1+\pi^2/3) \nonumber\\
&& + \pi/3(\lambda^2)(Y^3)(Z^2- 1)^2
\end{eqnarray}

where ,  $ Y =  Rf_{\pi} $,    $ Z = \frac{F}{ f_{\pi}} $ ,  $ f_{\pi}
= 93 $ Mev is  the pion decay constant and and we take $\lambda^2 = 42 $
corresponding to a sigma mass  of 850 Mev.

We can then calculate, $  E_{CS} $ at a given, R, which corresponds to a cell length, $ a = 2R $ and baryon density, $ n_B = \frac{1}{(2R)^3} $. We
then minimise the energy with respect to  Z(F) . This 
shows a trend that as the density goes up, the value of F increases, which indicates that the spontaneous breaking of chiral symmetry is enhanced as baryon density increases.

$  E_{CS} $, is the energy per soliton in the crystal, without any inter soliton interaction or canonical quantization which we attempt compute below

i) Realistically, the tensor interaction energy between the solitons in the chiral
limit  ( $ m_{\pi} = 0$) needs to be modulated by a factor  $ \exp (-m_{\pi)}
a)/(a^3) $ , where , $ a$, is the  cell length, $ a = 2R $ . In the
chiral limit the tensor interaction  From Fig. 1 \cite{Kleb} can be
normalised at a cell length $ a = 2R = 2.15 $ fm, which is equivalent  to
a cell volume  $ V = (n_b)^{-1} = 10 fm^3 $, and is found to be , $ \sim  -
70$ Mev. When we apply the pion mass correction, $\exp(-m_{\pi}a) $, to
this it comes down to - 15.4 Mev. We use this as the normalisation for
the inter soliton tensor interaction which is modulated by the factor,
 $\exp(-m_{\pi} a)/(a^3) $ as the cell length decreases.
The energy of tensor interaction  that follows is 

\begin{equation}
  E_T =  \exp (-m_{\pi} a)/(\frac{a}{2.15})^3 \cdot 70   Mev
\end{equation}

We must mention here that the above tensor interaction is an asymptotic form but we have persisted with it at separations of , $ a \sim 1 $  fm, where this assumption may not be valid. 

ii) 
The energy of canonical quantization for our composite
soliton is very similar to the skyrme soliton above. As stated before
the difference between the two, $  \Delta E_Q = E_2 - E_1$,\cite{Kleb}
gives us the contribution of the energy of canonical quantization.

 Including i) and ii) above, the total energy per baryon  for
 our case of a crystal of composite solitons with quark bound states is
 given by

\begin{equation}
  E_B    =  E_{CS} + \Delta E_Q +  E_T
\end{equation}

We note again that the crystal state above is a stable state only well above nuclear density.

\begin{figure}
  \includegraphics[width=\columnwidth]{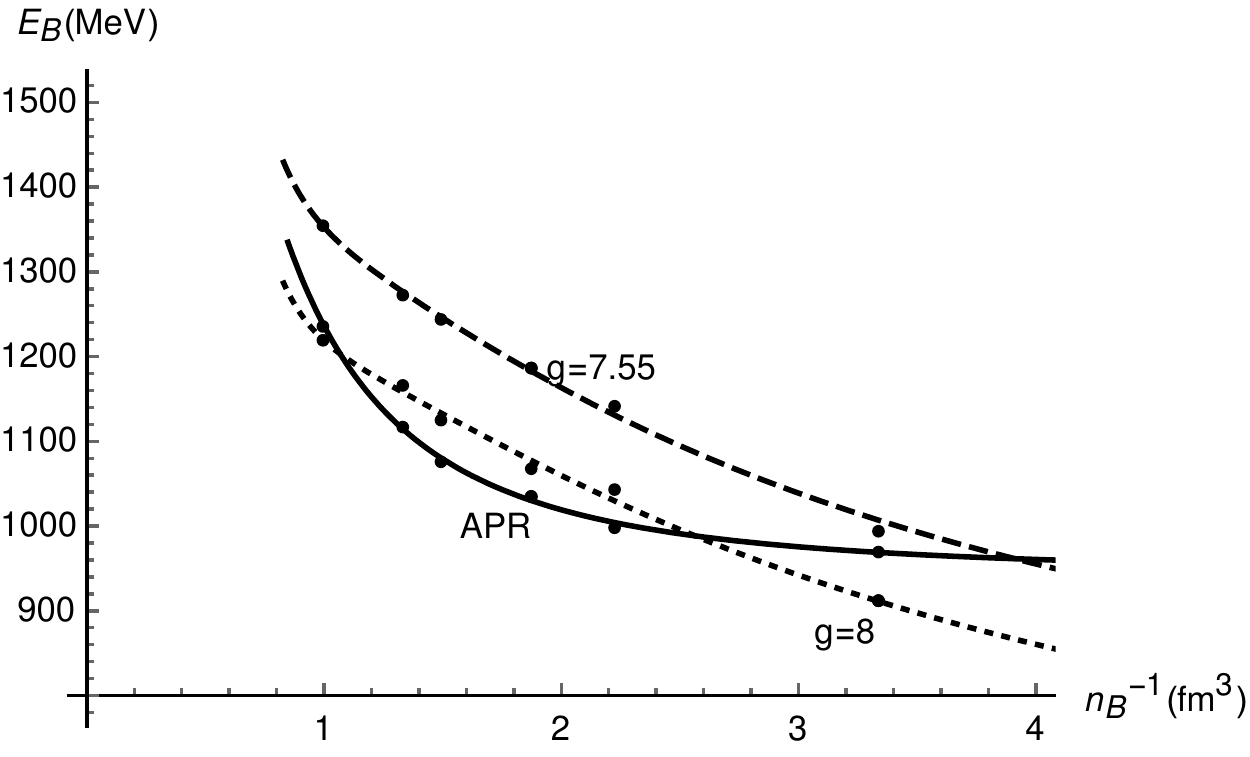}
\caption{Energy per baryon, $E_B$, for the APR EOS, and the quark soliton crystal EOS for, g = 7.55 and g= 8} 
 
\label{Fig4}
\end{figure}

\subsection{Remarks}

i) First, this is indeed a  new state of dense matter that has not been explored before. Whereas, conventional quark matter is a fermi
liquid in the presence of a stationary wave neutral pion condensate \cite{Soni1, Soni2}, in our composite nucleon soliton crystal state the quarks live in a 'topological'  crystal, in a frozen fully occupied , $ 0^+ $, band up to the threshold baryon density. This is also very different to conventional nuclear matter.
 
ii) It is important to note , as can be seen from (Section 6 ) in Ref.\cite{NPA}, that an exact solution brings down the  soliton energy by close to 25 percent.

iii) The inter soliton attractive  tensor interaction must be added. This can be a fairly large negative contribution has been added as indicated above from the pure skyrmion.

iv) In our estimate  ( see Fig, 4 ) we use the additive positive contribution of the isorotational energy of collective quantization. It is a moot question if we should have used the large negative contribution that follows from the projection procedure.

v) Next,there is the question of the zero point energy. In the pure
Skyrme soliton in the previous section each cell carries a skyrme soliton
with unit baryon number which is then localised and will carry zero point
energy. From the band structure in  ref \cite {BGS} we find that actually
the quark wave functions are not localised but are Bloch functions which occupy a filled band. The baryon number is carried by the quarks and not localised in a cell. Also, the pion and sigma fields are like a stationary wave condensate. Thus, in our model we do not have any zero point energy.

 vi) The APR equation works with point nucleons with the repulsive (
 hard core) interaction carried by, for example, the $ \omega $ meson
 interaction potential. Our repulsive ( hard core interaction) has
 different origin - the deep quark bound states.  Once we have finite size
 composite nucleons the space between nucleons is squeezed. The composite
 nucleons have a size which makes them overlap at  high density,
 generating a hard core.  Thus we  expect them to have a crystalline ground
 state. Such a state is not accessible for the APR nucleons which are
 point particles. Point like  nucleons would be difficult to localise  due
 their large zero point energy. 


vii)  Till now,  we have set our VEV's ( Sec. (III) to(VI)) for our soliton model with quark bound states in accordance with the single skyrmion configuration where the VEV for the pion and sigma fields are\cite {Soni1,BGS} 

\begin {equation}
<\sigma> = f_{\pi} Cos{\theta(r)},  <\vec\pi> = \hat r f_{\pi}Sin{\theta(r)}  
\end {equation}

    where, $\theta(r) = 0$,  at the cell boundary, r = R,  and 
		
		$ \theta(r\rightarrow 0)= -\pi$,  for the pion field to be well defined at the origin.

 But the constraint at the cell boundary is not required for the crystal which allows the pion field to
 be non zero at the cell boundary between the cells.  As the density is increased the pion field will goes up gradually at the cell boundary and goes to zero only at the centres of the solitons, thus  doubling the 'wavelength' of the pion field at very high density. This will reduce the energy per
baryon, as both the quark bound state eigenvalue and gradient energy can come down substantially.  This exercise was not carried out in ref \cite
{BGS} and will be presented in  a later work.

 Interestingly, though, very different from our solitonic crystal lattice, Pandharipande and Smith\cite{PS,BaymLN} do find a nucleon matter EOS that is a  crystalline solid of neutrons with a neutral pion condensate. Neutral pion condensation is also a  feature of our 3 flavour quark matter ground state\cite{Soni1}. It thus seems that a pion condensate is a uniform feature of both the quark  soliton state and the final quark matter state at high density.

viii) The Goldberger Trieiman relation which follows from PCAC introduces a renormalisation  factor for nucleons that ups the axial current coupling, $g_A $, from 1 to $\sim 1.36 $ (see section 6.5 in Baym \cite{BaymLN}).  The quarks in our model are similar to nucleons and  acquire their mass from the  spontaneous breaking of chiral symmetry which gives them a large constituent mass. We may thus expect a corresponding increase in the coupling, g.

ix)  We have assumed a simple cubic crystal till now and a corresponding baryon density, $ n_B \sim (1/2R)^3 $. A hexagonal close packed structure is also possible, which will have higher density for the same , R,  compared to the cubic structure. This can reduce the threshold density at which the quark matter transition occurs.

Furthermore, as in the previous sections, we have used the same  approximate soluble model parametrization of the sigma and pion fields. An exact solution will yield a lower value of the coupling, g, in turn increasing the value of, R ($ R \sim 1/g $ ), and reducing the density at which band gets to the continuum.

x)  Though, we have not included, (vii). (viii) and (ix), but used the reduction for the exact solution indicated in , (ii), above, we find  an EOS that is similar to the APR ( see Fig. 4) but somewhat above it. With a slight increment in  , $g \sim 8$, as suggested in (viii) we can recover an EOS that is close to the APR ( see Fig. 4). Recall, that the EOS for the relativistic crystalline (quark solitonic) nucleon state is good only well above nuclear density.

Our rough estimates should be viewed as a demonstration that at high baryon density an APR like EOS is possible  for  a crystal of composite solitons with quark bound states. As posted earlier we are engaged in an ongoing work in which we use a full solution for, $  E_{CS} $, including all the contributions listed above.


xi) The maximum mass for the neutron star for our model can then be taken
to be similar to that for the APR EOS. From the APR EOS \cite{APR} this the maximum mass at such central densities is $\sim 2.1 - 2.3 $ solar masses. We note that for the solid nucleon crystal model of Pandharipande and Smith\cite{PS} with a pion condensate the maximum mass is of the same order.


 Once the density hits a threshold, where the quarks are no longer
bound or frozen in the $0^+ $ band, we can transit into normal or conventional quark matter. As indicated earlier this happens roughly when $ n _B \sim 1/fm^3$. Once the barrier at this threshold is overcome, we expect that nuclear matter to make a sudden transition into  pion condensed quark matter. This is the point when the EOS becomes soft through a decompression. The sudden increase in density can mimic a collapse  generating a shock wave which ejects matter.


\section{ Energy release in Merger of neutron stars }


 The sudden phase transition from the relativistic crystalline ( quark soliton) state to  'free' quark matter will result in a contraction or the core. To illustrate this we calculate the pressure using the  APR nuclear matter EOS  from fig.1 in Ref. \cite{Soni2}  and the quark matter EOS from our Fig. 1. We note that the pressure in the APR EOS we use goes up sharply at high density. If the sudden phase transition to  'free' quark matter occurs at around, $n_B \sim 1/fm^3$, we find that the pressure in the nuclear phase at this density is , $ P \sim 600 Mev/fm^3$.  Fig 5 illustrates the transition from APR nuclear matter  to quark matter at this pressure at $n_B \sim 0.95/fm^3$. Other equations of state are softer as is the crystalline state in Fig. 4. This would move the APS curve to the right and reduce the pressure at which this density occurs.The continuous line tracks  the evolution as the system goes to higher density. Of course, the pressure, P, and energy per baryon, $E_B$, in the 'free' quark matter state at this density are much lower. 

\begin{figure}
	\includegraphics[width=\columnwidth]{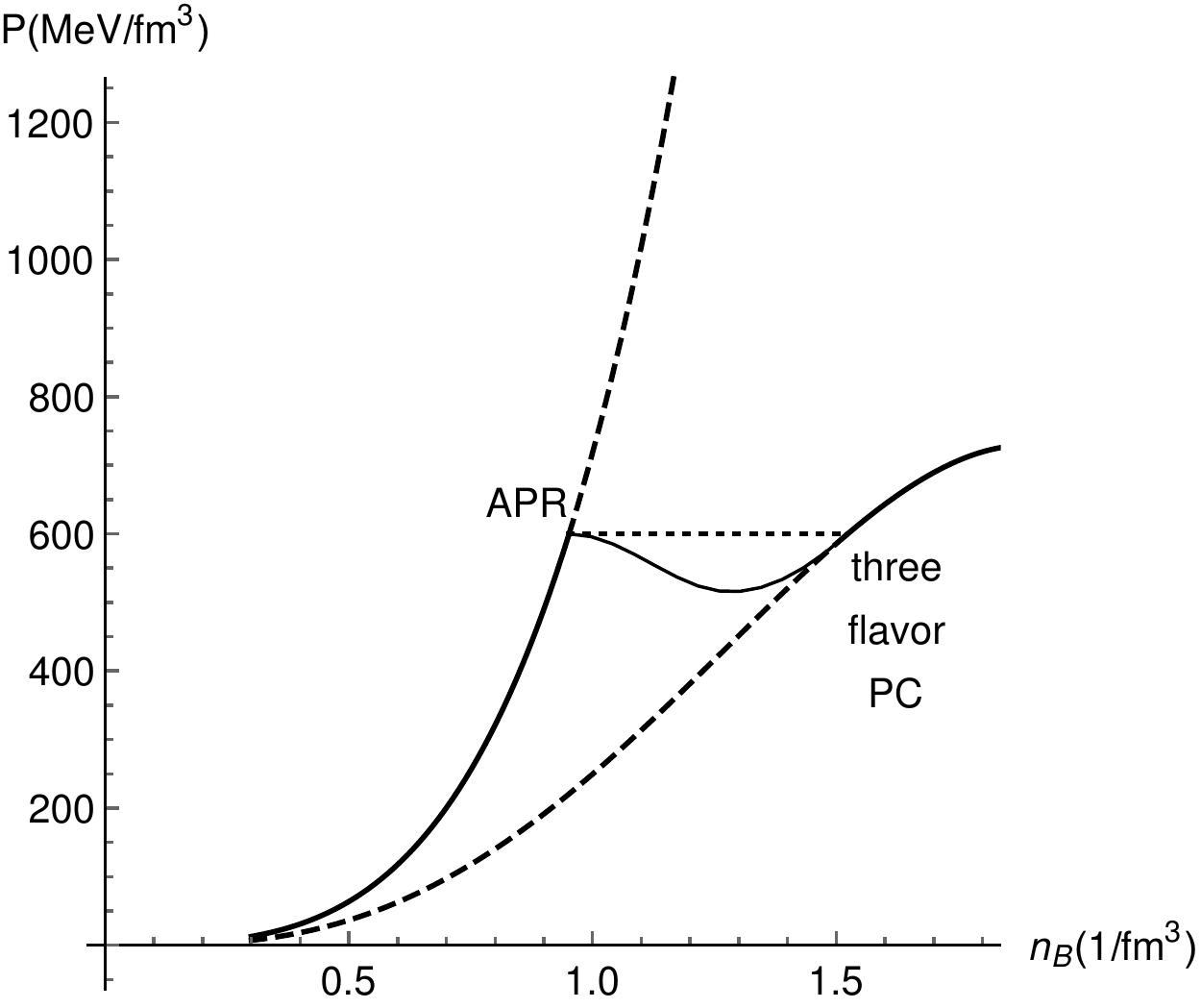}
\caption{Pressure, $ P ( Mev/{fm}^3 )$, vs  $n_B ( 1/{fm}^3 ) $,  for the APR EOS, and the pion condensed 3 flavour quark matter. Illustration of the transition that occurs at, $n_B ( 0.95/{fm}^3 )$, in the nuclear phase} 
 
\label{Fig5}
\end{figure}

Since we need to balance the pressure in both phases it is pertinent
to find the density at which the same pressure occurs  in the 'free'
quark matter' state. From Fig. 5 ,this is found to be,  $n_B \sim 1.5/fm^3$, and the corresponding, $ E_B \sim 1260 MeV$. Thus, as nuclear matter clears the threshold barrier set by  the soliton crystal, there will be a sudden contraction followed by a  consequent increase in density from, $n_{B1} \sim 0.95/fm^3$   to $n_{B2}\sim 1.5/fm^3$, as the system attains the same pressure. As pointed out before, if the EOS is softer than the APS the pressure and threshold density at which the quark matter transition occurs will come down.
 
 In a high mass neutron star or in the merger of 2 neutron stars such
 a major change in compressibility,  K, would cause a contraction of the core and  bring down the gravitational
 potential energy. A rough estimate( Newtonian) of the gravitational energy release is provided by considering a  neutron star of mass  $M$ whose potential
energy is,  $(3/5) G M^2/R$. Keeping the mass fixed we can write down the
energy difference as we change the density indicated above( see  Fig. 5)  from , $
\rho_1 \sim 1.6 \cdot 10^{15} $ gm/cc to $\rho_2 \sim 2.53\cdot 10^{15} $
gm/cc , for a uniform density star,
where, $R = (\frac{3M}{4 \pi\rho})^{1/3} $

   \begin {equation}
    \Delta E_G  =  (3/5) (G M^2) [ 1/( R_2) -  1/( R_1)  ]
        \end {equation}

 On substituting the the values of  a 2 solar mass star, $M \sim  4\cdot
 10^{33}$ gm  and the above density change for the corresponding radii, we can get
 a sudden release of gravitational energy of,  $ \Delta E_G  \sim 0.7 \cdot
 10^{53} $, ergs which can yield a matter ejecting shock wave.

\section{ Discussion  }

One significant difference with  most of the equations of state in the
literature and this work is that we have composite nucleons where the structure of the nucleon plays an
essential role in the transition from  nuclear matter to quark matter at
high density.  Working with nucleons that are chiral solitons with relativistic quark bound states we have presented evidence for  plausible new crystalline ground state for dense matter, at densities where nucleons overlap, to show that conventional quark matter does not occur till such density at which the quark bound states get compressed and merge with the continuum.  In this ground state it is topology and chiral quark interactions in the nucleon that determine the threshold transition density. This is in contrast to most other equations of state. Our motivation in Sec. VII is to make a rudimentary estimates for  the EOS of our composite solitons is to show that it can be potentially similar to the APR 98 EOS. As we have stated a more convincing calculation of the EOS which includes several improvements  will be presented separately.

These works indicate that strongly  bound quarks in the quark soliton
model of the nucleon translate into a 'hard core' interaction between
nucleons, resulting in an equation of state that provides even a stronger
nucleon nucleon repulsion than the hard core repulsion encountered in
the APR EOS. Such a ‘hard core’ interaction provides a  potential
barrier between the solitonic crystal phase and the normal quark matter phase. 
These considerations  modify the simple minded Maxwell construction above.


We would like to emphasize that the composite solitonic crystal state is a a new state of matter that is neither conventional quark matter nor conventional nuclear matter. One notable difference is that whereas  the  conventional quark matter state is a fermi liquid in the presence of a stationary wave neutral pion condensate, in our composite nucleon soliton crystal state, the quarks live  in a special, frozen, fully occupied, relativistic, $0^+$,  band up to the threshold baryon density. 

If the maximum mass of the neutron star for a particular EOS occurs below this density then we can say that neutron  stars exist entirely in the solitonic crystal  phase, and  become unstable even before the transition to quark matter. On the other hand if  the maximum mass for a particular EOS occurs above  this threshold central density we conclude that matter is unstable to transiting to quark matter even before the maximum mass of the star in the nucleonic phase. In any case, conventional high mass quark  matter stars  (for example, MIT bag) \cite{Soni1,BH} are unstable and very unlikely to ever have an EOS that can go up to a,  $ M_{max}\sim 2 $, solar mass.

For example, a neutron star based on the APR [A18 + dv +UIX] \cite{APR}
EOS has a maximum mass, $ M_{max} \sim 2.2 M_{solar} $, that occurs at
a central density slightly larger than $n_b \sim 1/fm^3 $. We expect the the star can be unstable and transit into quark matter even below this maximum mass. For a softer EOS, the maximum mass will come down, but the transition density to quark matter will go up and thus the star will most likely again become  unstable even before the maximum mass is reached. .

For a stiffer EOS, corresponding to  APR [A18 + UIX],  its maximum
mass is slightly higher , $ M_{max} \sim  2.3 M_{solar} $, than  that of
APR[A18 + dv +UIX], but at a central density which is lower than,  $n_B  \sim 1/fm^3 $. Thus  the star becomes unstable before the quark matter transition takes place. For the unrealistic case of an EOS  that is assumed to be incompressible at a density above 3 times nuclear saturation density ( see fig. 14 ref.\cite{APR}) we can expect a higher $ M_{max}$ but not greater than $ \sim 2.5 M_{solar} $. 

Our analysis indicates that  all stable neutron stars remain in the nucleonic soliton state and that their $ M_{max} $ will not exceed $ \sim 2.5 M_{solar} $. Secondly,  when the mass of a solitonic star (or coalesced stars) exceeds the maximum mass or the central density exceeds the threshold density, which ever happens earlier, there will be a sudden decompression ( or contraction ) transition  to an unstable quark matter state with an abrupt change in density. The hybrid crossover models \cite{BH} do not have such a density 'discontinuity' as in these models nuclear and quark matter can co exist with a smooth journey to the maximum mass. 

After the merger of two neutron stars the contraction of the core due to the sharp change in the compressibility of EOS at the threshold density of our model would result in a different post merger scenario, in contrast to hybrid crossover models where there is no such effect. This may be observable. This abrupt change in density could result in a shock wave and  matter ejection.  It is also possible that fast rotating binary mergers support a metastable, 'hypermassive', intermediate state.  As has been outlined in some earlier work  the passage to high mass stars is likely to produce magnetars \cite {mag} beyond a certain mass which can carry very high magnetic fields that can catalyse the formation of jets in this event.

In our model if the mass of the merger of two neutron stars  exceeds $ M_{max} \sim 2.5  M_{solar} $ then the final state will be a black hole. Given that the observed merger resulted in a final state that was, $ \sim 2.7 M_{solar} $, either way the merger will finally transit to a black hole. To conclude, in the alternative model we have presented, all neutron stars should have no regular `free'quark matter and that the transition for neutron stars with masses over the maximum mass, $ M_{max} $, or for central baryon density  larger than, $n_b \sim 1/fm^3 $, will become unstable and transit to `free'quark matter and then onto black holes .

Acknowledgement: We are happy to acknowledge discussions and a partial
collaboration with Dipankar Bhattachrya, Pawel Haensel and Mitja Rosina. The author thanks the centre for Theoretical Physics, Jamia Millia Islamia and ICTP, Trieste for hospitality.


\begin{thebibliography}{99}




\bibitem{shapiro10} Paul Demorest, Tim Pennucci, Scott Ransom, Mallory Roberts, Jason Hessels	Nature 467, (2010)1081-1083 
\bibitem {freire} J. Antoniadis P. C.C. Freire et al, arXiv:1304.6875 [astro-ph.HE]
\bibitem{Soni1} V. Soni and D. Bhattacharya,  Phys. Lett. B {\bf  643} (2006) 158.

\bibitem{BH}  G. Baym, et al, Reports on Progress in Physics 81(5)  (2018), arXiv:1707.04966v3 [astro-ph.HE] 2017

\bibitem{APR} A. Akmal, V. R. Pandharipande and D. G. Ravenhall, Phys. Rev. C {\bf 58} (1998) 1804.

\bibitem{Lattimer} J. M. Lattimer and M. Prakash, Astrophysical J. {\bf 550} (2001) 426;

\bibitem{bla1}T. Klahn, R. Łastowiecki,and D. Baschke, Phys. Rev.  D 88,(2013) 085001 
\bibitem{bla2} S. Benic et al. Astron.Astrophys. 577 (2015) A40
\bibitem{bla3}R. Lastowiecki, D. Blaschke, T. Fischer and T. Klähn arXiv:1503.04832v1 (2015)
\bibitem{hat1}K. Masuda, T. Hatsuda, and T. Takatsuka, Prog. Theor. Exp. Phys. (2012) 

\bibitem{hat2} G. Baym, T. Hatsuda, M. Tachibana, N. Yamamoto, J. Phys. G35, 104021 (2008).
\bibitem{fuk} K. Fukushima, Phys. Lett. B 591, 277 (2004)
\bibitem{baym} K. Yamazaki, T. Matsui, and G. Baym,   Nucl. Phys. A ·  933, (2015), 245

\bibitem{weise1}T. Hell, N. Kaiser, W. Weise, S. Schulte, B. R\"ottgers, 'New Constraints from Neutron Stars'- indico.cern.ch

\bibitem{weise2}S. Fiorilla, N. Kaiser, W. Weise, Nucl.Phys.A880, 65 (2012)


\bibitem{abbott1} B. P. Abbott et al, Phys. Rev. Lett. 119, 161101:1-18 (2017).
\bibitem{abbott2}B. P. Abbott et al, Astrophys. J. Letters 848, (2017)
\bibitem{Soni2} V. Soni and D. Bhattacharya, arXiv. Hep-ph/0504041 v2.
\bibitem{NPA} S. Kahana, G. Ripka and V. Soni, Nuclear Physics A {\bf 415} (1984) 351.
\bibitem{BB}  M. C. Birse and M. K. Banerjee, Phys. Lett. B  {\bf 136} (1984)  284.
\bibitem {dautry} F. Dautry,  and E. M. Nyman, 
   Nucl. Phys. A319 (1979) 323
	
\bibitem{kutschera+90} M.  Kutschera, 
W. Broniowski, and  Kotlorz, A. 1990, Nucl. Phys. A516, 566 

\bibitem{BGS} B. Banerjee, N. Glendenning and V. Soni Physics Letters B, Volume 

\bibitem{PS} V. R. Pandharipande and R. A. Smith, Nuclear Physics A237 (1975) 507-532
\bibitem{BaymLN}  G. Baym, Neutron Stars and the Properties of matter at High Density, Lecture Notes NBI and NORDITA (1977)

\bibitem{Kleb} I. Klebanov, Nuclear Physics B 262 (1985) 133-143 

\bibitem{mag} V. Soni and N. D.Haridass, MNRAS 425 (2):(2012)  1558-1566.






\end{thebibliography}
\end{document}